\documentstyle[psfig]{lamuphys}
\makeatletter
\let\chapter\hid@chapter
\makeatother
\begin{document}
\pagenumbering{arabic} 
\title{A Semiclassical Approach to Level Crossing in Supersymmetric
Quantum Mechanics}
\titlerunning{A Semiclassical Approach to Level Crossing in SUSY QM}

\author{J.~F.~Beacom
   \thanks{Speaker.
   Current address: Physics 161-33, Caltech, Pasadena, CA 91125, USA.\\
   {\tt beacom@citnp.caltech.edu}}\inst{}
   and A.~B.~Balantekin
   \thanks{\tt baha@nucth.physics.wisc.edu}\inst{}}
\institute{Department of Physics, University of Wisconsin,
           Madison, WI 53706, USA}

\maketitle


\begin{abstract}
Much use has been made of the techniques of supersymmetric quantum
mechanics (SUSY QM) for studying bound-state problems characterized by
a superpotential $\varphi(x)$.  Under the analytic continuation
$\varphi(x) \rightarrow i\varphi(x)$, a pair of superpartner
bound-state problems is transformed into a two-state level-crossing
problem in the continuum.  The description of matter-enhanced neutrino
flavor oscillations involves a level-crossing problem.  We treat this
with the techniques of supersymmetric quantum mechanics.  For the
benefit of those not familiar with neutrino oscillations and their
description, enough details are given to make the rest of the paper
understandable.  Many other level-crossing problems in physics are of
exactly the same form.  Particular attention is given to the fact that
different semiclassical techniques yield different results.  The best
result is obtained with a uniform approximation that explicitly
recognizes the supersymmetric nature of the system.
\end{abstract}


\section{SUSY QM and the Bound-State Problem}

Starting with a superpotential $\varphi(x)$, one can generate two
superpartner potentials
\begin{equation}
V_\pm = \varphi^2(x) \pm \frac{\hbar}{\sqrt{2 m}}\, \varphi'(x)\,.
\label{Vdef}
\end{equation}
For these two potentials the two corresponding Schr\"odinger equations
are
\begin{equation}
\left[
-\frac{\hbar^2}{2 m} \frac{\partial^2 }{\partial^2 x} + V_\pm
\right] \Psi_\pm(x)
= E \Psi_\pm(x)\,.
\label{Sdef}
\end{equation}
It can be shown that the eigenspectrum of the ``$+$'' system can be
obtained by shifting the quantum numbers $n$ of the ``$-$'' system by
$n \rightarrow n - 1$, with the ground state of the ``$-$'' system
discarded.  That is, the spectra of the two systems are identical
except for a single state.  In applications, this property is
exploited in the following way.  Given a potential $V(x)$, one
attempts to find a superpotential $\varphi(x)$ that will generate
$V(x)$ via Eq.~(\ref{Vdef}), with one or the other sign.  If this can
be done, one can immediately generate the superpartner potential by
choosing the opposite sign in Eq.~(\ref{Vdef}).  In some fortunate
circumstances, the equations of motion for the second system are much
easier to solve than the first.  See \cite{Schwabl} for more
introductory material, and \cite{Cooper} and references therein for
active areas of research.

The application of supersymmetric quantum mechanics to the solution of
bound-state problems has been extensively developed.  There has been
particular interest in semiclassical techniques.  A direct primitive
semiclassical (WKB) approach to the Schr\"odinger equation yields the
usual Bohr-Sommerfeld quantization condition:
\begin{equation}
\sqrt{2m} \int^{x_2}_{x_1} dx \sqrt{E - V(x)} =
\left(n + \frac{1}{2}\right)\hbar\pi\,,
\end{equation}
where $x_1$ and $x_2$ are the turning points (zeros of the integrand).
Since the superpartner potentials depend explicitly on $\hbar$, the
Schr\"odinger equations in Eq.~(\ref{Sdef}) have a different
dependence on $\hbar$ than the usual case.  A primitive semiclassical
(WKB) solution of Eq.~(\ref{Sdef}), which however explicitly
recognizes the supersymmetric nature of the system [\cite{Comtet}],
therefore yields a modified quantization condition:
\begin{equation}
\sqrt{2m} \int^{x_2}_{x_1} dx \sqrt{E - \phi^2(x)} = n\hbar\pi\,,
\end{equation}
where $x_1$ and $x_2$ are the turning points (zeros of the integrand).
This modified quantization condition is exact for many systems (see
\cite{catalog} for a catalog of results), but not all [\cite{Nieto}].
The WKB wave functions are singular at each turning point.  It is
possible to avoid this problem by using a uniform approximation -- one
that is valid for all $x$, including at the turning points.  The
approach is similar to WKB, but there is a special construction to
cancel the turning-point singularity.  A uniform semiclassical
solution [\cite{Susyform1}] of Eq.~(\ref{Sdef}) recovers the modified
quantization condition.  However, it gives a much better wave
function, which will be essential for the transition probability
derived below.


\section{Introduction to Neutrino Oscillations}


\subsection{Vacuum Neutrino Oscillations}

It is possible that the flavor and mass eigenstates of neutrinos are
not identical.  Throughout this paper, we consider mixing between only
two flavors, electron and muon.  Then a general state can be written
in the flavor basis:
\begin{equation}
|\nu(t)\rangle = \Psi_e(t) |\nu_e\rangle + \Psi_\mu(t) |\nu_\mu\rangle
\end{equation}
or the mass basis:
\begin{equation}
|\nu(t)\rangle = \Psi_1(t) |\nu_1\rangle + \Psi_2(t) |\nu_2\rangle\,.
\end{equation}
The amplitudes are defined as
\begin{displaymath}
\Psi_e\:(\Psi_\mu) = {\rm amplitude\ for\ the\ neutrino\ to\ have\ flavor\ }
e\:(\mu)
\end{displaymath}
\begin{displaymath}
\Psi_1\:(\Psi_2) = {\rm amplitude\ for\ the\ neutrino\ to\ have\ mass\ }
m_1\:(m_2)\,.
\end{displaymath}
The amplitudes are taken to be time-dependent, and the kets to be
time-independent.  For two flavors, the flavor and mass bases must be
related by a simple rotation.  This rotation is taken to be between
the amplitudes, with the kets held fixed, and is given by:
\begin{equation}
\left[\begin{array}{cc} \Psi_1(t) \\ \\ \Psi_2(t) \end{array}\right]
=
\left[\begin{array}{cc}
\cos{\theta_v} & -\sin{\theta_v} \\ \\
\sin{\theta_v} & \cos{\theta_v}
\end{array}\right]
\left[\begin{array}{cc} \Psi_e(t) \\ \\ \Psi_{\mu}(t) \end{array}\right]\,,
\end{equation}
where $\theta_v$ is the vacuum mixing angle.

In the mass basis, the Schr\"odinger equation is
\begin{equation}
i\hbar \frac{\partial}{\partial t}
\left[\begin{array}{cc} \Psi_1(t) \\ \\ \Psi_2(t) \end{array}\right] =
H_{\rm mass}
\left[\begin{array}{cc} \Psi_1(t) \\ \\ \Psi_2(t) \end{array}\right]\,,
\end{equation}
where
\begin{equation}
H_{\rm mass} =
\left[\begin{array}{cc}
E_1 & 0 \\ \\
0 & E_2
\end{array}\right]\,,
\end{equation}
and is diagonal by definition.  Since the neutrino masses $m_1$ and
$m_2$ are presumed small, we make an ultrarelativistic expansion
(using $c = 1$ units here and below).  Defining $E$ to be the common
energy, and defining the mass-squared difference as
\begin{equation}
\delta m^2 = m_2^2 - m_1^2\,,
\end{equation}
one can show that $H_{\rm mass}$ is given by
\begin{equation}
H_{\rm mass} =
\left(E + \frac{m_1^2 + m_2^2}{4E}\right)
\left[\begin{array}{cc}
1 & \phantom{+}0 \\ \\
0 & \phantom{+}1
\end{array}\right]
+ 
\frac{\delta m^2}{4E}
\left[\begin{array}{cc}
-1 & \phantom{+}0 \\ \\
\phantom{+}0 & \phantom{+}1
\end{array}\right]\,.
\end{equation}
Below, the term proportional to the identity matrix will be dropped.

Now we change to the flavor basis, using the rotation matrix above
that relates the two bases.
In the flavor basis, the Schr\"odinger equation is
\begin{equation}
i\hbar \frac{\partial}{\partial t}
\left[\begin{array}{cc} \Psi_e(t) \\ \\ \Psi_\mu(t) \end{array}\right] =
H_{\rm flavor}
\left[\begin{array}{cc} \Psi_e(t) \\ \\ \Psi_\mu(t) \end{array}\right]\,,
\label{SE_flav}
\end{equation}
where
\begin{equation}
H_{\rm flavor} =
\frac{\delta m^2}{4E}
\left[\begin{array}{cc}
-\cos{2\theta_v} & \sin{2\theta_v} \\ \\
\sin{2\theta_v} & \cos{2\theta_v}
\end{array}\right]\,.
\label{Hflav0}
\end{equation}
Using the rotation matrix, the amplitude to be of the electron type at
a time t is
\begin{equation}
\Psi_e(t) = \cos{\theta_v} \Psi_1(t) + \sin{\theta_v} \Psi_2(t)\,.
\end{equation}
The time evolution of the mass amplitudes is trivial, so that just a
phase relates the amplitudes at a point $t$ to those at the point $t =
0$.  The mass amplitudes at $t = 0$ can be expressed in terms of the
flavor amplitudes at $t = 0$ by use of the rotation matrix.  Taking as
initial conditions $\Psi_e(0) = 1$, $\Psi_\mu(0) = 0$ (an
electron-type neutrino produced at $t = 0$),
\begin{equation}
\Psi_e(t)
= \cos^2{\theta_v} 
\exp\left(+i \frac{\delta m^2}{4E}\frac{t}{\hbar}\right) 
+ \sin^2{\theta_v} 
\exp\left(-i \frac{\delta m^2}{4E}\frac{t}{\hbar}\right)\,.
\end{equation}
The probability for the neutrino to be of the electron type can then
be shown to be
\begin{equation}
P(\nu_e \rightarrow \nu_e)(t) = |\Psi_e(t)|^2 = 
1 - \sin^2{2\theta_v} \sin^2{\left(\pi t/L_{\rm osc}\right)}\,,
\end{equation}
where the oscillation length is $L_{\rm osc} = 4\pi E\hbar/\delta
m^2$, so called since this is the separation between extrema in the
survival probability.  Since the initial flavor state was not a
stationary state of the Hamiltonian, the probability to be of either
flavor oscillates.  Note that if either the vacuum mixing angle or the
mass-squared splitting is small, the effects of the oscillations are
minimal.  If either the source or detector has a finite size of order
$L_{\rm osc}$ or greater, then after averaging over that region,
\begin{equation}
\langle P(\nu_e \rightarrow \nu_e) \rangle_{\rm avg} = 
1 - \frac{1}{2}\sin^2{2\theta_v}\,,
\end{equation}
which is independent of energy.


\subsection{Matter-Enhanced Neutrino Oscillations}

In this section we assume the vacuum oscillations discussed above.
Here we consider that the neutrino is traveling through a medium with
a varying density of electrons, e.g., the sun.  We consider the
scattering
\begin{equation}
\nu_x + e^- \rightarrow \nu_x + e^-\,,
\end{equation}
where $\nu_x$ indicates either flavor.  In such reactions, a neutrino
of a given flavor enters and a neutrino of the same flavor leaves.
Electron neutrinos undergo both charged- and neutral-current reactions
with the electrons, but muon neutrinos undergo only neutral-current
scattering.  For elastic forward scattering, the coherence of the
neutrino ``beam'' can be maintained.  The effect of the medium is to
modify the dispersion relation (refractive index) of the neutrino.
Equivalently, one can say that the neutrino masses are modified in the
medium, and that the modification is different for electron and muon
neutrinos.  Below, we neglect a term proportional to the identity, and
consider only the difference between the electron and muon neutrinos.
Then the Hamiltonian in Eq.(\ref{Hflav0}) becomes
\begin{equation}
H_{\rm flavor} =
\frac{\delta m^2}{4E}
\left[\begin{array}{cc}
\zeta(t) -\cos{2\theta_v} & \sin{2\theta_v} \\ \\
\sin{2\theta_v} & - \zeta(t) + \cos{2\theta_v}
\end{array}\right]\,,
\label{Hflav1}
\end{equation}
where $\zeta(t)$ is related to the electron number density $N_e(t)$ by
\begin{equation}
\zeta(t) = \frac{2\sqrt{2} G_F N_e(t)}{\delta m^2/E}\,,
\end{equation}
and $G_F$ is the Fermi coupling constant.

When $\zeta(t) \rightarrow 0$, the vacuum mixing case is recovered.
When $\zeta(t) \rightarrow \infty$, the Hamiltonian is also
considerably simplified.  However, when $\zeta(t) = \cos{2\theta_v}$,
the effect of the off-diagonal coupling is maximal.  This is known as
the resonance point.  In a medium with a varying density, such as the
sun, it is possible for an electron neutrino to be produced at a
density above the resonance density.  Then during its passage out of
the sun it will pass through the resonance density.  That has profound
consequences for the probability that it emerges in vacuum as an
electron neutrino.

The flavor-basis Hamiltonian can be instantaneously diagonalized with
a rotation matrix, the angle of which is called the matter angle
$\theta(t)$.  The bases are related as:
\begin{equation}
\left[\begin{array}{cc} \Psi_1(t) \\ \\ \Psi_2(t) \end{array}\right]
=
\left[\begin{array}{cc}
\cos{\theta(t)} & -\sin{\theta(t)} \\ \\
\sin{\theta(t)} & \cos{\theta(t)}
\end{array}\right]
\left[\begin{array}{cc} \Psi_e(t) \\ \\ \Psi_{\mu}(t) \end{array}\right]\,.
\end{equation}
As the density varies, so does the matter angle.  At high densities,
the matter angle $\theta \rightarrow \pi/2$; at resonance, $\theta =
\pi/4$; and at low densities, $\theta \rightarrow \theta_v$, the
vacuum mixing angle (this is explained in \cite{Susynu}).

\begin{figure}
\psrotatefirst
\psfig{file=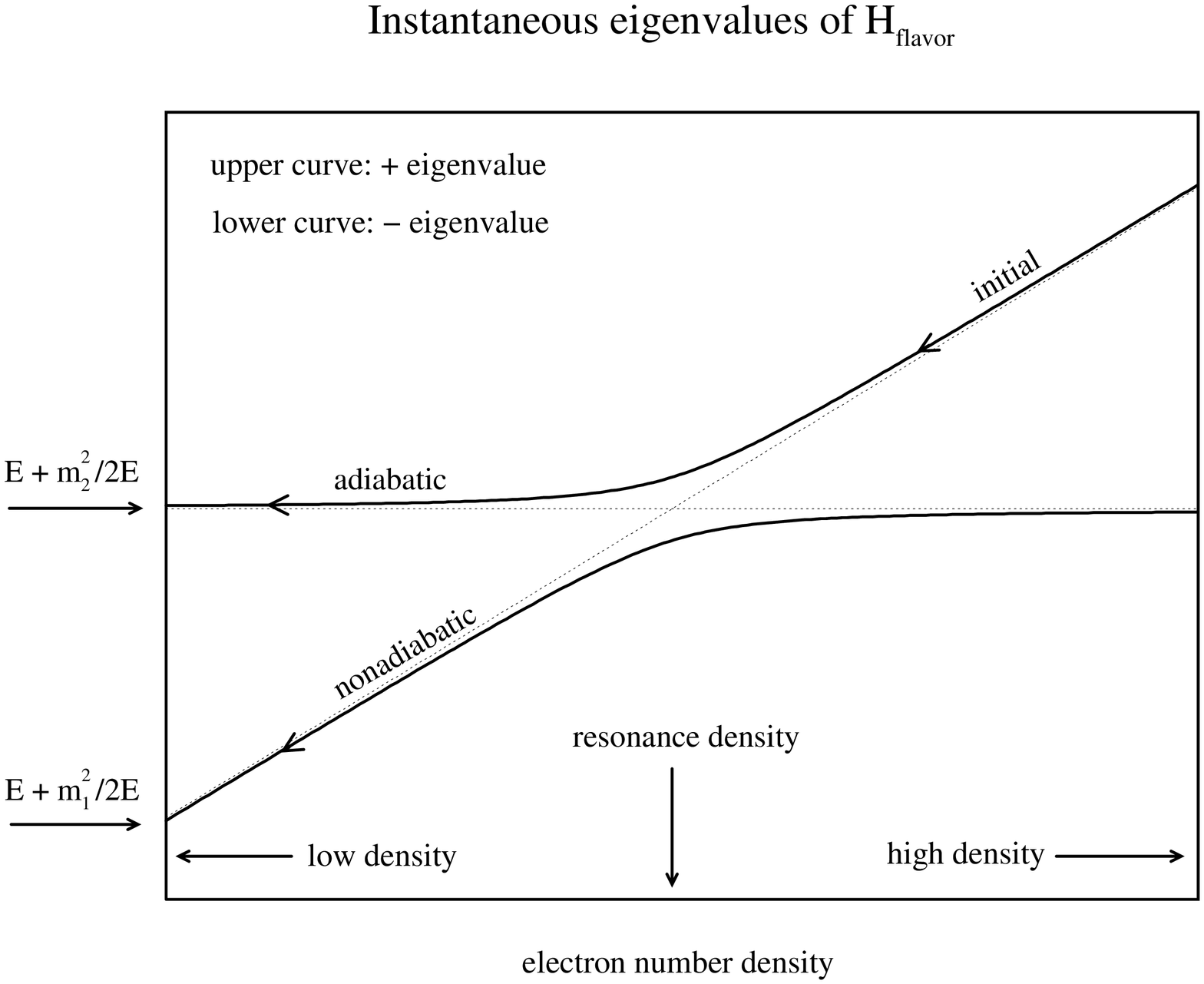,width=11.7cm,angle=-90}
\caption{A schematic illustration of how the instantaneous energy
eigenvalues of $H_{\rm flavor}$ vary as a function of electron number
density.  The details depend upon the mixing parameters, the energy,
and the density profile.  Beyond the left edge of the figure, the
density is assumed to vanish; the eigenvalues are constant in vacuum
and have the values indicated.  In the figure, the initial state is
assumed to be a mass = $m_2$ eigenstate.  The final state may be the
same eigenstate (adiabatic case) or the other eigenstate (nonadiabatic
case).  In the figure, the initial and final densities appear to be
equally far from the resonance; that is not true in general.}
\label{crossing}
\end{figure}

The instantaneous eigenvalues of the flavor-basis Hamiltonian are
(reintroducing the term proportional to the identity matrix):
\begin{equation}
E +
\frac{m_1^2 + m_2^2 + \delta m^2 \zeta(t)}{4E} \mp
\frac{\delta m^2}{4E} \sqrt{\sin^2{2\theta_v} + 
(\zeta(t) - \cos{2\theta_v})^2}\,.
\end{equation}
If the density changes slowly (adiabatically), then the instantaneous
diagonalization will be almost exact and the mass eigenstates will be
almost stationary states.  (Note that the Hamiltonian is exactly
diagonal in the mass basis for any constant density.)  If the density
changes quickly (nonadiabatically), then the neutrino can ``hop'' from
one eigenstate to another in the neighborhood of the resonance (where
the splitting between mass eigenvalues is at a minimum).  The
probability of this occurring is called the hopping probability
$P_{\rm hop}$.  This occurs proportionally to the extent that the
instantaneous diagonalization fails.  The instantaneous eigenvalues of
$H_{\rm flavor}$ as a function of density are shown schematically in
Fig.~\ref{crossing}.

Consider an electron neutrino created at a high density and a time
$t=0$.  Since the matter angle will be near $\pi/2$, then $\Psi_e(0)
\approx \Psi_2(0)$.  If the density changes slowly, the neutrino will
remain in this eigenstate.  Then, if the vacuum mixing angle is fairly
small, then $\Psi_2(t) \approx \Psi_\mu(t)$ in vacuum.  In this case,
the electron neutrino created in the solar core will emerge as a muon
neutrino, and it will be missed by experiments counting electron
neutrinos from the sun.  On the other hand, if the density changes
rapidly the neutrino may emerge in the other instantaneous eigenstate.
Then, if the vacuum mixing angle is fairly small, then $\Psi_1(t)
\approx \Psi_e(t)$ in vacuum.  In this case, the neutrino emerges as
an electron neutrino after all.

In a later section, this problem is reformulated so that these extreme
limits of the matter and vacuum angle are not necessary.  In general
then, whether the neutrino emerges from the sun as the electron or
muon type is a complicated function of the mixing parameters, the
energy, and the form of the density profile.  The probability of it
emerging as the original flavor is called the survival probability
$P(\nu_e \rightarrow \nu_e)$; the general form is given in
Eq.~(\ref{pnu}).

For further introductory reading on neutrino oscillations, see
\cite{Neutrino-Intro}.  For a recent review of neutrino astrophysics
(including oscillations), see \cite{Astronu}.


\section{Supersymmetric Character of the Level-Crossing Problem}

The Hamiltonian of Eq.~(\ref{Hflav1}) is a typical form for a
level-crossing problem.  Away from the resonance point, the
off-diagonal elements may be neglected.  The diagonal elements
approximate the eigenvalues, and are allowed to vary.  If the
off-diagonal elements were exactly zero, the eigenvalue trajectories
would cross (become equal) at the resonance point.  This crossing will
be avoided if the off-diagonal elements have any nonzero value, no
matter how small.  In the most general level-crossing problem, the
off-diagonal elements would also be allowed to vary.  Here they are
taken to be constant.  Nevertheless, that is not a large practical
restriction.  Approximating the off-diagonal elements as constant is
reasonable as the resonance region (the only region in which they
contribute significantly) is usually very narrow.  The explicit
representation of the supersymmetry in Eq.~(\ref{Hflav1}) is given in
\cite{Susyform2}.

Before proceeding further, we switch to working with dimensionless
quantities.  We define a length scale
\begin{equation}
L = \frac{\hbar\lambda}{\delta m^2 /4 E}\,,
\label{L}
\end{equation}
and use this to define $x = t/L$.  In Section~1, $x$ was used to
denote a generic coordinate (with dimensions of length).  Here and
below, $x$ is dimensionless.  Since we will be making a semiclassical
expansion, we need to be able to keep track of formal powers of
$\hbar$.  For each $\hbar$ in the problem, we write $\lambda$ and
consider $\lambda$ to be formally small; this is equivalent to saying
that the length $L$ is small.  We will make expansions in powers of
$\lambda$, truncating the higher orders.  At the end of the
calculation, we will set $\lambda = 1$.  For notational convenience,
we redefine the flavor-basis Hamiltonian as follows:
\begin{equation}
i\lambda \frac{\partial}{\partial x}
\left[\begin{array}{cc} \Psi_e(x) \\ \\ \Psi_{\mu}(x) \end{array}\right]
= H_{\rm flavor}(x)
\left[\begin{array}{cc} \Psi_e(x) \\ \\ \Psi_{\mu}(x) \end{array}\right]\,,
\end{equation}
where
\begin{equation}
H_{\rm flavor}(x) =
\left[\begin{array}{cc}
\eta\varphi(x) & \sqrt{\Lambda} \\ \\
\sqrt{\Lambda} & -\eta\varphi(x)
\end{array}\right]\,.
\label{Hflav2}
\end{equation}
We have defined
\begin{equation}
\eta\varphi(x) = \zeta(x) - \cos{2\theta_v}
\label{varphi}
\end{equation}
and
\begin{equation}
\Lambda = \sin^2{2\theta_v}\,.
\label{Lambda}
\end{equation}
(The definition of $\Lambda$ was misprinted in \cite{Susynu}).
The scaled electron density is
\begin{equation}
\zeta(x) = \frac{2 \sqrt{2}\ G_F E N_e(x)}{\delta m^2}\,.
\end{equation}
Note that there are notation changes from previous related works
[\cite{Susyform1}, \cite{Susyform2}, \cite{Susyform3}]; here we have
made $\Lambda$ and $\varphi$ dimensionless.  The factor $\eta$ (taken
to be $\pm 1$), is introduced above to control the analytic behavior
of the function $\varphi(x)$ in the complex plane, as explained in
\cite{Susynu}.  In the expressions with $\varphi^2$ below, we drop
$\eta^2 = 1$.

The coupled first-order equations of Eq.~(\ref{Hflav2}) can be
decoupled to yield
\begin{equation}
-\lambda^2\frac{\partial^2 \Psi_e(x)}{\partial x^2} -
\left[\Lambda + \varphi^2(x) + i\lambda\eta\varphi'(x)\right]
\Psi_e(x) = 0
\label{SE_e}
\end{equation}
and
\begin{equation}
-\lambda^2\frac{\partial^2 \Psi_{\mu}(x)}{\partial x^2} -
\left[\Lambda + \varphi^2(x) - i\lambda\eta\varphi'(x)\right]
\Psi_{\mu}(x) = 0\,,
\label{SE_m}
\end{equation}
where $\varphi(x)$ and $\Lambda$ are defined in Eqs.~(\ref{varphi})
and (\ref{Lambda}).  Such a simple decoupling is not possible in the
mass basis.  Eqs.~(\ref{SE_e}) and (\ref{SE_m}) are explicitly of the
supersymmetric form.  This form follows directly from the (rather
general) form of the level-crossing Hamiltonian above.  The two levels
are superpartners.

These Schr\"odinger-like equations are similar to those for
non-relativistic particles in the presence of a complex barrier, and
for convenience we use the language of wave mechanics to describe
them.  In particular, to the extent that we can ignore the imaginary
terms in the potential, these correspond to particles above a barrier
(since $\Lambda > 0$).  There are two caveats regarding discussing
this as a barrier penetration problem.  First, that our boundary
conditions do not correspond to the usual picture of incident,
reflected, and transmitted waves; in general, there are waves moving
in each direction on each side of the barrier.  The boundary
conditions make it a level-crossing problem instead of a barrier
problem.  (The condition of a pure electron neutrino at the initial
point requires $\Psi_e = 1$, $\Psi_\mu = 0$ there.)  Second, the pure
imaginary terms in the potentials play an extremely important role
here, even in the asymptotic regions.  These terms are needed not only
to represent nonadiabatic transitions, but also to allow the local
matter angle to change.  Since our problem is a level-crossing
problem, the quantity of interest is not a reflection or transmission
coefficient, but rather $P(\nu_e \rightarrow \nu_e) = |\Psi_e(x
\rightarrow \infty)|^2$, the probability of the neutrino being of the
electron type far from the source.


\section{Semiclassical Solution of the Equations of Motion}


\subsection{General Form of the Solution}

In general, the survival probability has terms which depend on the
source and detector positions.  In this paper, these interference
terms are considered to be averaged away by the finite source and
detector sizes (or by a varying distance between them, such as that
due to the motion of the earth).  The remaining term below must also
be considered to have been appropriately averaged.

\begin{figure}[t]
\centerline{\psfig{file=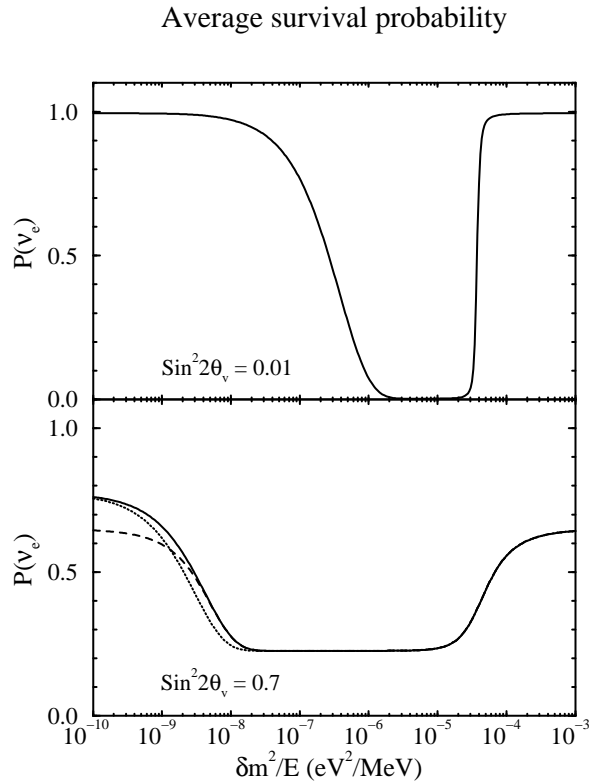,width=7.8cm}}
\caption{The electron neutrino survival probability vs.\ the
mass-squared difference parameter for two different vacuum mixing
angles.  The solid line is given by the method of
\protect\cite{Susynu}.  The dashed line is the exact (numerical)
result.  The dotted line is the linear Landau-Zener result.  In the
top figure, the lines are indistinguishable.  An exponential density
with parameters chosen to approximate the sun was used
[\protect\cite{Bahcall}].  The region leftward of the lower left
corner of the trough is the nonadiabatic region.}
\label{pitplot}
\end{figure}

The general form of the survival probability is
\begin{equation}
P(\nu_e \rightarrow \nu_e) = 
\frac{1}{2}
\left[1 + (1 - 2P_{\rm hop})\cos{2\theta_i} \cos{2\theta_v}\right]\,.
\label{pnu}
\end{equation}
The two matter angle terms account for the rotations between the mass
and flavor bases at the initial and final (in vacuum) points.  (The
matter angle thus ranges from $\pi/2$ at infinite density to
$\theta_v$ in vacuum.  At the resonance, $\theta = \pi/4$.)  $P_{\rm
hop}$ is the probability of hopping from one mass eigenstate to the
other during the transit through the resonance region.  This result
for the survival probability is completely general.  The heart of the
problem is to determine $P_{\rm hop}$.  Various semiclassical results
for $P_{\rm hop}$ are discussed below.  The variation of the averaged
survival probability with $\delta m^2/E$ for two choices of the vacuum
mixing angle is shown in Fig.~\ref{pitplot}.

The general shape of the survival probability vs $\delta m^2/E$ is
that of a pit.  High- and low-energy electron neutrinos have a
relatively high probability to remain so, whereas medium-energy
electron neutrinos have a relatively low probability to remain so (and
hence a relatively high probability to transform into muon neutrinos).
The solar neutrino detectors are primarily sensitive to electron
neutrinos.  The energy-dependent suppression of electron neutrinos as
in Fig.~\ref{pitplot} explains the solar neutrino data well.

\subsection{Primitive Semiclassical Solution}

A primitive semiclassical (WKB) solution can be made to
Eqs.~(\ref{SE_e}) and (\ref{SE_m}).  It can be shown that this is
exactly equivalent to the adiabatic solution.  The adiabatic solution
is obtained when the Hamiltonian changes so slowly that the
instantaneous diagonalization always holds with high accuracy.  That
is, there are no transitions between mass eigenstates, so $P_{\rm hop}
= 0$ by definition.  This result is not shown in Fig.~\ref{pitplot}.
If it were, it would match the exact result in all but the left-hand
rise of the pit.  Where the other curves rise up, the WKB result would
remain flat, at the level of the bottom of the pit.

In the bound-state problem, even the primitive semiclassical solution
gave an excellent answer for the quantization condition.  As noted, in
the neutrino problem this approximation does not predict $P_{hop}$.
The problem comes from the well-known singularity of the WKB wave
functions near the turning points.  In this problem, the turning
points are in the complex plane, near the resonance point (which is on
the real axis).  In the bound-state problem, the turning point
singularity is not as crucial since the quantization condition only
counts nodes between the turning points.  In this problem, we need the
wave function itself to be accurate, since its modulus squared
determines the survival probability.


\subsection{Landau-Zener Solution}

Far from the resonance, the propagation is adiabatic, and can be
trivially solved for any density profile.  A reasonable approach to
simplifying the problem is to approximate the density profile as
linear in the resonance region.  This is the basis of the Landau-Zener
solution.  In the flavor-basis Hamiltonian of Eq.~(\ref{Hflav2}), this
gives the usual Landau-Zener setup: linear variation of the diagonal
elements, and constant off-diagonal coupling.  In order to make
Fig.~\ref{crossing} more general, the variation is shown versus the
density itself (instead of $x$).

The solution is obtained from the decoupled equations (\ref{SE_e}) and
(\ref{SE_m}).  With $\zeta(x) \sim x$, the differential equations can
be reduced to the defining equation for the parabolic cylinder (Weber)
functions.  Given the initial conditions $\Psi_e = 1, \Psi_\mu = 0$,
the solution for $\Psi_e(x)$ at any point is straightforward.
(Actually, the boundary conditions require some care since the linear
density eventually becomes negative, which is unphysical.)  From the
asymptotic form appropriate far after the resonance, one can extract
the hopping probability $P_{\rm hop}$ from the the survival
probability $P(\nu_e \rightarrow \nu_e) = |\Psi_e(x \rightarrow
+\infty)|^2$.  The result is
\begin{equation}
P_{\rm hop} = 
\exp\left(-\pi\frac{\delta m^2}{4 E\hbar}
\frac{\sin^2{2\theta_v}}{\cos{2\theta_v}}
\left|\frac{\dot\zeta(t)}{\zeta(t)}\right|^{-1}_{\rm res}\right)\,.
\end{equation}

Because the Landau-Zener solution is based on the exact solution for
the linear density, there are no turning-point singularities as with
the WKB solution.  Thus the approximate wave function and hence
$P_{\rm hop}$ are fairly reasonable.  However, there is some
inaccuracy due to the fact that a general density does vary more than
linearly through the resonance region.  The fact that the Landau-Zener
result has the right general behavior but is not very accurate can be
seen in Fig.~\ref{pitplot}.


\subsection{Uniform Semiclassical Solution}

In this section we outline a uniform semiclassical solution to the
problem [\cite{Susynu}].  An arbitrary monotonic density profile is
allowed, as are nearly arbitrary mixing parameters.  We start with
Eq.~(\ref{SE_e}):
\begin{equation}
-\lambda^2\frac{\partial^2 \Psi_e(x)}{\partial x^2} -
\left[\Lambda + \varphi^2(x) + i\lambda\eta\varphi'(x)\right]\Psi_e(x)
= 0\,.
\end{equation}
There is no known exact solution of this for arbitrary $\zeta(x)$.
Recall that $\lambda$ is a formally small perturbation parameter (we
set $\lambda = 1$ later), and that $\eta = \pm 1$.  Compare this to
\begin{equation}
-\lambda^2\,\frac{\partial^2 U(S)}{\partial S^2} -
\left[\Omega + S^2 \pm i\lambda\eta\right]U(S) = 0\,,
\end{equation}
which is solvable in terms of parabolic cylinder functions.
Identification of the turning points of each of these will be crucial.
At lowest order in $\lambda$, they are the points for which $\Lambda +
\varphi^2(x) = 0$ and $\Omega + S^2 = 0$, respectively.  The turning
points are complex conjugate pairs.

Suppose we could find a change of variables $S = S(x)$.  Using this,
we could deform the simple parabolic ``barrier'' of the second
equation into the more complicated ``barrier'' shape of the first.
Then the solutions of the general case could be expressed in terms of
the solutions of the simple case.  Such a formal solution can be
written as
\begin{equation}
\Psi_e(x) = \frac{1}{\sqrt{S'(x)}} U(S(x))\,.
\end{equation}

Starting with this formal solution, and the two Schr\"odinger
equations above for $\Psi_e(x)$ and $U(S)$, one can determine a
differential equation for $S = S(x)$.  Unsurprisingly, that is a
nonlinear equation, and at least as difficult as the original
Schr\"odinger equation.  However, this equation can be profitably
subjected to a semiclassical solution.  We expand as
\begin{equation}
S(x) = S_0(x) + \lambda S_1(x) \dots\,,
\end{equation}
and truncate at second order.  Separating the various orders in
$\lambda$ determines a series of equations.  The solution of the ${\cal
O}(\lambda)$ equation determines $S_0(x)$:
\begin{equation}
\int^{S_0(x)}_{i\sqrt\Omega} {dS_0 \sqrt{\Omega + S_0^2}} =
\int^{x}_{x_0}
{dt \sqrt{\Lambda + \varphi^2(x)}}
\label{S0}
\end{equation}
and also $\Omega$:
\begin{equation}
\Omega = \frac{2i}{\pi}
\int^{x_0^*}_{x_0}
{dx \sqrt{\Lambda +\varphi^2(x)}}\,,
\end{equation}
where $x_0$ and $x_0^*$ are the turning points (zeros of the
integrand).  The ${\cal O}(\lambda^1)$ equation determines $S_1(x)$:
\begin{eqnarray}
S_1(x) & = & \frac{i\eta}{2\sqrt{\Omega + S_0^2(x)}} \nonumber \\
& \times & \left\{
\ln{\left[\frac{\varphi + \sqrt{\Lambda + \varphi^2(x)}}
{\sqrt\Lambda}\right]} +
\ln{\left[\frac{\sqrt{\Omega}}{S_0(x) + \sqrt{\Omega + S_0^2(x)}}\right]}
\right\}
\end{eqnarray}

The change of variables $S = S(x)$ is actually a mapping of the
complex $x$-plane to the complex $S$-plane.  In order to avoid
spurious branch cut discrepancies, the mapping must be chosen to not
fold or flip the plane.  This is accomplished with some appropriate
choices of signs.  Further, the turning points in the $x$-plane must
be mapped onto the turning points in the $S$-plane.  That demand fixes
the definition of $\Omega$ to be the one given above.

The approximate (but uniformly valid) solution is
\begin{equation}
\Psi_e(x) \approx
\left[
\frac{\Omega + S_0^2(x)}{\Lambda + \varphi^2(x)}\right]^{1/4}
U(S_0(x) + \lambda S_1(x))
\end{equation}
At the turning points, the denominator vanishes, just as in the WKB
solution.  However, because of the matching of turning points in the
mapping, the numerator vanishes at the same points.  That cancels the
singularity and gives an excellent approximation to the wave function
at all points, including near the turning points (and hence near the
resonance, which is on the real axis near the turning points).

As noted above, the functions $U(S)$ are parabolic cylinder functions.
Using the defining equations for $S_0(x)$ and $S_1(x)$ given above,
one can solve for $\Psi_e(x)$ for any $x$.  The implicit definition of
$S_0(x)$ makes it analytically solvable only for large $|x|$ and large
$|S_0|$, which holds far from the resonance.  However, both the
initial and final point can be taken to be far from the resonance.  At
each point, the general solution can be written in terms of two
independent parabolic cylinder functions with arbitrary coefficients.
After taking the asymptotic forms (meaning well away from the
resonance), one has:
\begin{eqnarray}
\Psi_e(x \rightarrow -\infty)
& & \\
& & \hspace{-2cm} = \:
C_- \cos\theta(x) \exp\left(+i I_p(x,x_i)/\lambda\right)
+ D_- \sin\theta(x) \exp\left(-i I_p(x,x_i)/\lambda\right)\,,
\nonumber
\end{eqnarray}
\begin{eqnarray}
\Psi_e(x \rightarrow +\infty)
& & \\
& & \hspace{-2cm} = \:
C_+ \sin\theta(x) \exp\left(-i I_p(x,x_i)/\lambda\right)
+ D_+ \cos\theta(x) \exp\left(+i I_p(x,x_i)/\lambda\right)\nonumber\\
& & \hspace{-2cm} = \:
\left[
c_1 \sin\theta_i \exp\left(+i {\rm Re} I_p(x_i,x_0)/\lambda\right) +
c_2 \cos\theta_i \exp\left(-i {\rm Re} I_p(x_i,x_0)/\lambda\right)
\right]\nonumber\\
& & \times
\sin\theta(x) \exp\left(-i {\rm Re} I_p(x,x_0)/\lambda\right)
\nonumber \\
& & \hspace{-2cm} + \:
\left[
c_1^* \cos\theta_i \exp\left(-i {\rm Re} I_p(x_i,x_0)/\lambda\right) -
c_2^* \sin\theta_i \exp\left(+i {\rm Re} I_p(x_i,x_0)/\lambda\right)
\right]
\nonumber \\
& & \times
\cos\theta(x) \exp\left(+i {\rm Re} I_p(x,x_0)/\lambda\right)\,,
\nonumber
\end{eqnarray}
where
\begin{equation}
I_p(x,x_i) = \int^{x}_{x_i}
{dx \sqrt{\Lambda + \varphi^2(x)}}\,.
\end{equation}
Above, $x$ is a general point, $x_i$ is the initial point, and $x_0$
is a turning point.  Now one must solve the problem of connecting the
coefficients $C_-$ and $D_-$ (known from the initial conditions) to
the coefficients $C_+$ and $D_+$ (or equivalently $c_1$ and $c_2$).
Tedious algebra reveals the solution to the connection problem to be:
\begin{eqnarray}
c_1 & = & -\frac{\Gamma(-\nu)}{\sqrt{2\pi}}
\Omega^{-i\Omega/2 + \eta/2}
\left(\frac{e^{-i\pi/2}}{2}\right)^\nu
\frac{e^{-3i\pi/4}}{\sqrt{2}} \exp\left(+\frac{i\Omega}{2}\right)
2i\sin\left(\nu\pi\right)
\\
c_2 & = & e^{-i\nu\pi}\,.
\end{eqnarray}
Comparison of $|\Psi_e(x \rightarrow + \infty)|^2$ above to the
general form
\begin{equation}
P(\nu_e \rightarrow \nu_e) = 
\frac{1}{2}
\left[1 + (1 - 2P_{\rm hop})\cos{2\theta_i} \cos{2\theta_v}\right]
\end{equation}
reveals that:
\begin{equation}
1 - P_{\rm hop} = |c_1|^2 = 1 - e^{-\Omega\pi}\,,
\end{equation}
\begin{equation}
P_{\rm hop} = |c_2|^2 = e^{-\Omega\pi}\,,
\end{equation}
where
\begin{equation}
\Omega = \frac{2i}{\pi} \left(\frac{\delta m^2}{4E\hbar}\right)
\int^{t_0^*}_{t_0}
{dt \sqrt{\sin^2{2\theta_v} + (\zeta(t) - \cos{2\theta_v})^2}}\,.
\end{equation}
The dimensions have been restored and $\Lambda$ and $\varphi$ replaced
by their definitions; recall that $\zeta(t) \sim$ density and $t_0,
t_0^*$ are the turning points (zeros of the integrand).  For a linear
density, the Landau-Zener result is recovered, i.e.,  
\begin{equation}
{\rm linear\ density:\ }
\Omega = \frac{\delta m^2}{4 E\hbar}
\frac{\sin^2{2\theta_v}}{\cos{2\theta_v}}
\left|\frac{\dot\zeta(t)}{\zeta(t)}\right|^{-1}_{\rm res}\,.
\end{equation}
However, any
arbitrary monotonic density may be used.
For example, for an exponential density,
\begin{equation}
{\rm exponential\ density:\ }
\Omega = \frac{\delta m^2}{4 E\hbar}
(1 - \cos{2\theta_v})
\left|\frac{\dot\zeta(t)}{\zeta(t)}\right|^{-1}\,,
\end{equation}
which is the leading term in the exact result.  That the approximation
is rather good over a wide variety of parameters can be seen in
Fig.~\ref{pitplot}.


\section{Conclusions}

In a semiclassical approximation, one always finds that the wave
functions have terms like
\begin{equation}
\Psi \sim \exp\left(\left(\ \ \right)/\hbar\right)\,,
\end{equation}
where $(\  \ )$ indicates a phase integral and various numerical factors.
There is an essential singularity in $\Psi$ in the formal limit $\hbar
\rightarrow 0$.  That is, $\Psi$ does not have a well-defined value in
this limit.  Different approximations (i.e., different ways of
limiting $\hbar \rightarrow 0$) therefore give different results.  Not
all semiclassical approximations are equivalent; other criteria must
be used to decide which approach to use.  In this level-crossing
problem, the best results come from a uniform approximation (valid for
all $x$, including the turning points) that explicitly recognizes the
supersymmetric nature of the potential.  Given the success of similar
approximations in the bound-state problem, this is perhaps to be
expected.

Even the uniform supersymmetric approximation eventually breaks down.
In the original problem (in the figure above, based on the exponential
density), we considered only the two turning points closest to the
real axis.  In general, there are more, further out in the complex
plane.  The distance of the the turning points from the real axis
scales like $|\zeta/\dot\zeta|$ (and for the exponential density,
integer multiples of this).  The analog potential (based on the linear
density), only has two turning points.  This mismatch in turning-point
topologies eventually makes the mapping function multivalued, at which
point the approximation breaks down.  The problem is worst when the
length scale $E \hbar/\delta m^2$ becomes large and ``sees'' further
into the complex plane.

This is an interesting practical problem, in part because there is
good evidence for matter-enhanced neutrino oscillations.  The
technique given here is applicable for any monotonic density profile
(and not just in the sun), and for a large range of mixing parameters
[\cite{Susynu}].  In addition, the techniques here should also be
useful in a variety of the continuum level-crossing problems that are
ubiquitous in quantum mechanics.  As noted, certain kinds of two-level
systems are automatically of the supersymmetric form.  The techniques
of semiclassical SUSY QM for the bound-state problem can be carried
over, as shown in our example.


\section*{Acknowledgments}

This work was supported in part by the U.S. National Science
Foundation Grant No.\ PHY-9605140 at the University of Wisconsin, and
in part by the University of Wisconsin Research Committee with funds
granted by the Wisconsin Alumni Research Foundation.



\end{document}